\documentclass[prl,showpacs,twocolumn,floatfix]{revtex4}

\usepackage{amsmath,amssymb}
\usepackage{graphicx}

\begin{document}

\newcommand{\Z}{Z_{\rm eff}}
\newcommand{\Zr}{\Z^{(\rm res)}}
\newcommand{\eps}{\varepsilon}

\title{Positron-molecule annihilation by capture into vibrational Feshbach
resonances of infrared-active modes}

\author{G. F. Gribakin}
\email[E-mail address: ]{g.gribakin@am.qub.ac.uk}
\author{C. M. R. Lee}
\email[E-mail address: ]{c.lee@qub.ac.uk}

\affiliation{Department of Applied Mathematics and Theoretical Physics,
Queen's University, Belfast BT7 1NN, Northern Ireland, UK}

\begin{abstract}
Enhanced positron annihilation on polyatomic molecules is a long-standing and
complex problem. We report the results of calculations of resonant positron
annihilation on methyl halides. A free parameter of our theory is the positron
binding energy. A comparison with energy-resolved annihilation rates measured
for CH$_3$F, CH$_3$Cl, CH$_3$Br [Barnes et al. Phys. Rev. A {\bf 74}, 012706
(2006)] shows good agreement and yields estimates of the binding energies.
\end{abstract}


\maketitle


In this paper we calculate the positron-molecule annihilation rate due to
resonant capture of positrons by infrared-active vibrational modes, and observe
good agreement with recent experimental data for methyl halides \cite{BYS06}.

When a fast positron interacts with matter, it undergoes a quick succession
of ionizing and other inelastic collisions, and slows down to eV or thermal
energies before annihilation. The low-energy positron annihilation rate in a
gas with number density $n$ is usually parametrized as
\begin{equation}\label{eq:Zeff}
\lambda \equiv \sigma_a vn =\pi r_0^2c n\Z,
\end{equation}
where $\sigma_a$ is the annihilation cross section, $v$ is the positron
velocity, $c$ is the speed of light, $r_0$ is the classical electron radius,
and $\Z$ is an {\em effective number of electrons} per gas
atom or molecule, that contribute to annihilation \cite{Fraser,QED}.
Originally, $\Z$ was introduced in expectation that the annihilation rate
would be in proportion to the number of target electrons, $Z$.
However, early experiments \cite{Deutsch:51,Paul:63,Tao:65} and later
systematic studies \cite{Heyland:82,Surko:88,Iwata:95} found that for many
polyatomic molecules $\Z$ exceeded $Z$ by orders of magnitude.
It also showed strong chemical sensitivity and rapid growth with molecular
size (see review \cite{SGB05}).

Explanations of high molecular $\Z$ were sought in terms
of positron virtual or weakly bound states \cite{GS64},
resonances \cite{Smith:70,Ivanov}, long-lived vibrationally
excited positron-molecule complexes \cite{Surko:88}, and virtual Ps 
formation \cite{LW97}. At the same time, annihilation calculations
which neglected molecular vibrations, failed to reproduce ``anomalous''
$\Z$ for polyatomics \cite{Gianturco,Lima}, but gave evidence that
$\Z$ depend on the molecular geometry \cite{NG03}.

These efforts highlight the fact that positron-molecule annihilation is a
complex problem. Nevertheless, a theory developed in Refs.
\cite{Gr00,Gr01,GG04} provides a framework for analyzing this phenomenon.
There are two basic mechanisms of positron annihilation, direct and resonant.
The direct mechanism applies to both atoms and molecules and involves
annihilation of an incident positron ``in flight''. Its contribution is
enhanced when a low-lying virtual or weakly bound positron state is present,
leading to $\Z$ up to $10^3$ for room-temperature positrons \cite{Gr00,Gr01}.

Resonant annihilation occurs for molecules capable of binding the positron.
To be captured into a bound state, the positron energy must be absorbed by
a vibrational excitation of the positron-molecule complex.
This gives rise to a vibrational Feshbach resonance (VFR) at the incident
positron energy $\eps = E_\nu +\eps _0 $, where  $E_\nu $ is the vibrational
excitation energy, and $\eps_0<0$ is the positron bound state energy.
The positron bound in the VFR can annihilate (or undergo detachment).
The probability of annihilation is proportional to the resonance lifetime.

For non-monoenergetic positrons and closely spaced resonances, their
contribution to $\Z$ is proportional to the vibrational level density at
$E\approx \eps -\eps _0$ \cite{Gr00,Gr01}. If the positron VFR were due to
excitation of fundamentals alone, this density would be proportional to the
number of modes. Experimental $\Z$ show much faster increase (e.g., $\Z=3500$,
$11\,300$, and $37\,800$ for C$_3$H$_8$, C$_4$H$_{10}$, and C$_5$H$_{12}$,
respectively). This means that positron attachment involves excitation of
overtones and combination vibrations. Large $\Z$ are then related to high
total vibrational spectrum densities in the polyatomics.

The important role of vibrations was recently verified by measuring the
energy dependence of $\Z$ at sub-eV energies with a high-resolution positron
beam \cite{Gilb:02,BGS03}. These experiments uncovered peaks in $\Z$, whose
energies corresponded to those of molecular vibrational modes. In particular,
for all alkanes larger than methane, $\Z$ displayed a prominent C--H maximum.
Its downshift from the C--H mode energy (0.37 eV) provided a measure of the
positron binding energy \cite{bind}. Observation of such peaks means that
excited fundamentals act as vibrational doorway states \cite{GG04}, leading
to multimode vibrations through intramolecular vibrational relaxation
(IVR).


Therefore, to compute $\Z$ for polyatomics, one must account for strong
electron-positron correlations and positron binding, the interaction between
positronic and vibrational degrees of freedom and intramolecular vibrational
mixing. This makes {\em ab initio} calculations of high molecular $\Z$ very
difficult. However, as we show below, for small polyatomics some basic
features of resonant annihilation can be tested by relatively simple
calculations.

The resonant part of the annihilation cross section can be written using the
Breit-Wigner formalism \cite{Gr01,GG04,Landau},
\begin{equation}\label{eq:siga}
\sigma _a=\frac{\pi}{k^2}\sum _\nu \frac{g_\nu \Gamma_\nu^a \Gamma_\nu ^e}
{(\eps -E_\nu - \eps _0)^2+\Gamma_\nu^2/4},
\end{equation}
where $\Gamma _\nu^a$, $\Gamma_\nu^e$, and $\Gamma_\nu$ are the annihilation,
elastic and total widths of $\nu $th resonance, $g_\nu$ is its degeneracy,
and $k$ is the positron momentum (atomic units are used).
The annihilation width is proportional to the electron density at the
positron in the positron bound state, $\rho _{ep}$,
\begin{equation}\label{eq:Gam_a}
\Gamma _\nu^a=\pi r_0^2c\rho _{ep}.
\end{equation}
From Eqs. (\ref{eq:Zeff}) and (\ref{eq:siga}), the resonant $\Z$ is given by
\begin{equation}\label{eq:Zeffres}
\Zr =\frac{\pi }{k}\rho_{ep}\sum_\nu\frac{g_\nu \Gamma_\nu^e}
{(\eps -E_\nu - \eps _0)^2+\Gamma_\nu^2/4}.
\end{equation}
We will now use this equation to calculate the contribution of infrared-active
modes to $\Zr$.

Consider a compact polyatomic molecule that can bind the positron with a small
binding energy $|\eps _0|\equiv \kappa^2/2 \ll$ 1 eV.
The wavefunction of the bound positron is very diffuse and behaves as
$\varphi _0=Ar^{-1}e^{-\kappa r}$ outside the molecule. Since large 
distances dominate, the normalization constant is given by
$A\simeq (\kappa /2\pi )^{1/2}$ \cite{DO88}.

Suppose that the vibrational modes in this small-sized polyatomic
are not mixed with overtones or combination vibrations. Given the smallness
of the binding energy, the vibrational excitation energies of the
positron-molecule complex should be close to the fundamental frequencies
$\omega _\nu $ of the neutral molecule. In this case the sum
in Eq. (\ref{eq:Zeffres}) is over the modes $\nu $, and
$E_\nu \approx \omega _\nu $. Some (or even all) of these modes can be
infrared active. The positron capture into such excited states is mediated
by the long-range dipole coupling. This allows one to calculate their
contribution to $\Zr$.

Consider a positron with momentum ${\bf k}$ incident on the molecule in the
vibrational ground state $\Phi_0 ({\bf R})$, where ${\bf R}$ represents all
the molecular coordinates. If $k^2/2\approx \omega _\nu +\eps _0$, the
positron can be captured in the VFR, where it is bound to the molecule in a
vibrationally excited state $\Phi _\nu ({\bf R})$. The corresponding
width $\Gamma_\nu^e $ can be found from
\begin{equation}\label{eq:Gam_e}
\Gamma_\nu^e = 2 \pi \int |A_{\nu {\bf k}}|^2
\delta (k^2/2-\omega_\nu -\varepsilon_0) \frac{d^3k}{(2\pi)^3},
\end{equation}
where $A_{\nu {\bf k}}$ is the capture amplitude. We calculate it
by using a method similar to the Born-dipole approximation
\cite{La80,MS05}, as
\begin{eqnarray}
A_{\nu {\bf k}} &=& \int \varphi _0({\bf r})\Phi_\nu ^\ast ({\bf R})
\frac{\hat{\bf d}\cdot {\bf r} }{r^3}e^{i{\bf k}\cdot{\bf r}}\Phi_0 ({\bf R})
\,d{\bf r}d{\bf R} \nonumber \\
&=& \frac{4\pi i}{3}\,\frac{{\bf d}_\nu \cdot {\bf k}}{\sqrt{2\pi\kappa}}
\, _2F_1\left(\frac{1}{2},1;\frac{5}{2};-\frac{k ^2}{\kappa^2 }\right),
\label{eq:Anuk}
\end{eqnarray}
where $\hat{\bf d}$ is the dipole moment operator for the molecule,
${\bf d}_\nu =\langle \Phi _\nu |\hat{\bf d}|\Phi _0\rangle $, and
$_2F_1$ is the hypergeometric function \cite{hyper}.
Substitution of Eq. (\ref{eq:Anuk}) into Eq. (\ref{eq:Gam_e}) gives
\begin{equation}\label{eq:Gam_eh}
\Gamma _\nu ^e =\frac{16\omega _\nu d_\nu ^2}{27}\, h(\xi ),
\end{equation}
where $h(\xi )=\xi ^{3/2}(1-\xi )^{-1/2}
\left[ _2F_1\left(\frac{1}{2},1;\frac{5}{2};-\xi /(1-\xi )\right)
\right]^2 $
is a dimensionless function of $\xi =1+\eps _0/\omega _\nu$, such that
$\xi (0)=\xi (1)=0$, and $h_{\rm max}\approx 0.75$ at $\xi \approx 0.89$.

Equation (\ref{eq:Gam_eh}) shows that the elastic
width of a positron VFR for an infrared active mode is basically determined by
its frequency $\omega _\nu $ and transition dipole amplitude $d_\nu $,
known from infrared absorption measurements \cite{BC82}.

For weakly bound positron states the density $\rho _{ep}$ is a linear
function of $\kappa $ \cite{Gr01}. It can be estimated as
\begin{equation}\label{eq:rhoep}
\rho_{ep}= (F/2\pi )\kappa ,
\end{equation}
with $F\approx 0.66$ \cite{Gr01}. The same constant characterizes the
contribution of direct annihilation, $\Z^{\rm(dir)}\simeq F/(\kappa^2+k^2)$
\cite{Gr01}. It is enhanced at small positron momenta by the presence of a
weakly-bound (or virtual) state \cite{GS64,DF93}.


In a recent paper \cite{BYS06} measurements of $\Z$ for CH$_3$Cl and
CH$_3$Br using a cold trap-based positron beam, have been reported.
The energy dependence of $\Z$ for these molecules (and CH$_3$F measured
earlier \cite{BGS03}) shows peaks close to the vibrational mode energies.
This points to an important contribution of resonant annihilation in all three
molecules, although the maximum $\Z$ value for CH$_3$F (250) is much lower
than those for CH$_3$Cl and CH$_3$Br (1600 and 2000, respectively).

These molecules have $C_{3v}$ symmetry, and all six of their
vibrational modes are infrared active (see Table \ref{tab:CH3Cl} for CH$_3$Cl).
Methyl halides are also relatively small, which means that IVR may not
take place \cite{SM83}. This makes them ideal for application of our theory.
Equations (\ref{eq:Zeffres}), (\ref{eq:Gam_eh}), and (\ref{eq:rhoep}) allow
one to calculate the contribution of {\em all} VFR to $\Zr$, and the only free
parameter of the theory, i.e., the positron binding energy, can be chosen by
comparison with experimental $\Z$.

\begin{table}[ht!]
\caption{Characteristics of the vibrational modes of CH$_3$Cl.}
\label{tab:CH3Cl}
\begin{ruledtabular}
\begin{tabular}{cccrcc}
Mode & Symmetry & $g _\nu $ & $\omega_\nu $\footnote{Mode energies
$\omega_\nu $ and dipole amplitudes $d_\nu$ from Ref. \protect\cite{BC82}.}
& $d_\nu$ & $\omega_\nu d_\nu ^2$ \\
 & & & (meV) & (a.u.) & (a.u.) \\
\hline
$\nu_1$ & a$_1$ & 1 & 363 & $0.0191$ & $4.87\times 10^{-6}$ \\
$\nu_2$ & a$_1$ & 1 & 168 & $0.0176$ & $1.91\times 10^{-6}$ \\
$\nu_3$ & a$_1$ & 1 &  91 & $0.0442$ & $6.52\times 10^{-6}$ \\
$\nu_4$ & e & 2 & 373 & $0.0099$ & $1.34\times 10^{-6}$ \\
$\nu_5$ & e & 2 & 180 & $0.0162$ & $1.74\times 10^{-6}$ \\
$\nu_6$ & e & 2 & 126 & $0.0111$ & $5.66\times 10^{-7}$
\end{tabular}
\end{ruledtabular}
\end{table}

In order to do this, $\Z^{(\rm res)}$ from Eq.~(\ref{eq:Zeffres}) must be
averaged over the energy distribution of
the positron beam \cite{BGS03}. The latter can be modelled by a combination
of the Gaussian distribution in the longitudinal direction
($z$) and Maxwellian distribution in the transversal direction ($\perp $).
The corresponding probability density of the total positron energy,
$\eps _\perp +\eps _z$, is
\begin{equation*}\label{eq:f}
f_\epsilon (\eps _\perp ,\eps _z)=\frac{1}{k_BT_\perp \sqrt{2\pi \sigma^2}}
\exp \left[ -\frac{\eps _\perp }{k_BT_\perp}
-\frac{(\eps _z -\epsilon )^2}{2\sigma^2}\right],
\end{equation*}
where $k_B$ is the Boltzmann constant, $T_\perp $ is an effective transversal
temperature of the beam, $\epsilon $ is the mean longitudinal energy of the
positrons, as measured by the retarding potential analyzer, and
$\sigma=\delta _z/\sqrt{8\ln 2}$, $\delta _z$ being the full width at
half-maximum. The values of $k_BT_\perp $ and $\delta _z$ are taken from
experiment to be 25 meV.

The averaging, $\bar Z_{\rm eff}^{\rm (res)}(\epsilon)=\int \Zr
f_\epsilon (\eps _\perp ,\eps _z) d\eps _\perp d\eps _z $, can be done
analytically, since the
widths of the resonances, $\Gamma _\nu =\Gamma _\nu^e+\Gamma _\nu^a$, are
small compared to the energy spread of the positron beam. Indeed, values from
the last column of Table \ref{tab:CH3Cl} show that the elastic widths of the
VFR are less then $0.1$~meV. Typical annihilation widths are even smaller.
For example, for a binding energy of 10~meV ($\kappa =0.027$~a.u.),
Eqs. (\ref{eq:Gam_a}) and (\ref{eq:rhoep}) yield
$\Gamma _\nu ^a =3\times 10^{-9}~{\rm a.u.}= 0.1~\mu {\rm eV}$. These
estimates also show that $\Gamma _\nu \approx \Gamma _\nu^e$, i.e., that
the total width of the resonance is dominated by its elastic width.

Hence, to integrate over $\eps _\perp $ and $\eps _z$ we
replace the Breit-Wigner profiles in Eq. (\ref{eq:Zeffres}) by
$\delta $-functions, and obtain
\begin{equation}\label{eq:Zeff_fin}
\bar Z _{\rm eff}^{(\rm res)}(\epsilon )=
2\pi^2\rho_{ep}\sum_\nu \frac{g_\nu \Gamma _\nu^e}{k_\nu \Gamma _\nu}
\Delta (\epsilon -\eps _\nu),
\end{equation}
where $\eps _\nu = k_\nu ^2/2=\omega _\nu +\eps _0$ is the resonance energy, and
\begin{eqnarray}
\Delta (E)&=&\frac{1}{k_BT_\perp }
\exp \left[ \frac{\sigma ^2}{2(k_BT_\perp )^2}\right]
\exp \left(\frac{E}{k_BT_\perp }\right)\nonumber \\
&\times & \left\{ 1+\Phi \left[ -\frac{1}{\sqrt{2}}
\left(\frac{E}{\sigma }+\frac{\sigma}{k_BT_\perp }\right)
\right]\right\}, \label{eq:Delta}
\end{eqnarray}
with $\Phi (x)$ being the standard error function.

\begin{figure}[!ht]
\includegraphics*[width=7cm]{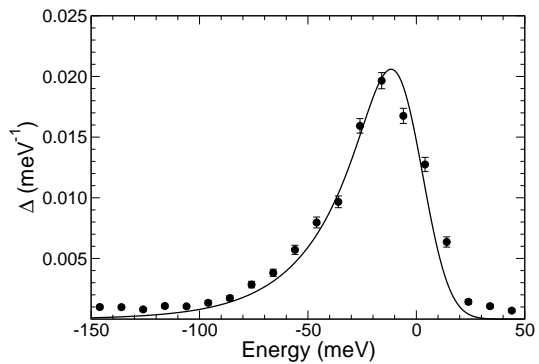}
\caption{Comparison of the resonance shape function $\Delta (E)$,
for $k_BT_\perp =\delta_z=25$~meV (curve) with the measured C--H peak
in propane (circles) \cite{Gilb:02,BGS03}. For comparison, experimental
$\Z$ has been scaled vertically and shifted horizontally.}
\label{fig:Delta}
\end{figure}

The function $\Delta (E)$ is a convolution of the $\delta $-function with
the positron energy distribution. It describes the appearance of a narrow
resonance when measured with the trap-based positron beam, and is shown in 
Fig.~\ref{fig:Delta}. Due to the transversal energy component, its maximum
is downshifted by 12 meV from the true resonance position. The shape of
$\Delta (E)$ is also markedly asymmetric, with an extended low energy tail.
It agrees well with those of the observed C--H peaks \cite{Gilb:02,BGS03}.
Note that the positron energy distribution was taken into account in
experiment by assuming a 16 meV difference between the positron total and
longitudinal energies \cite{BGS03}.

\begin{figure}[!ht]
\includegraphics[width=7cm]{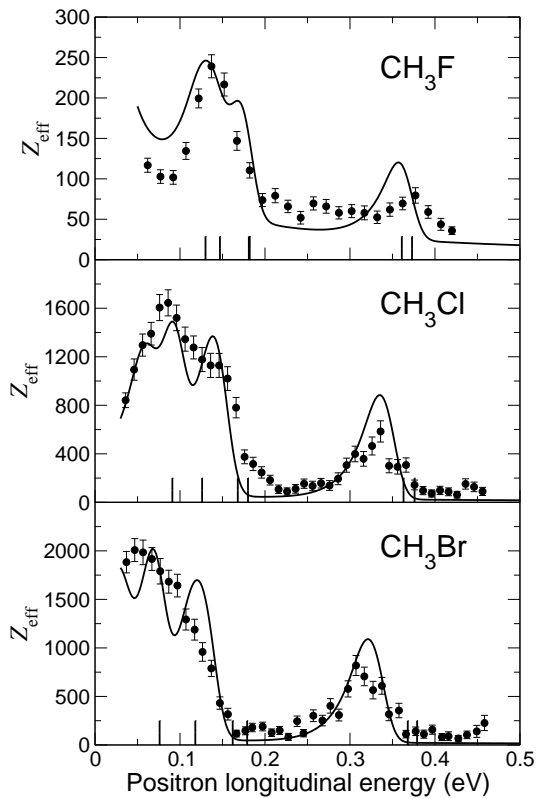}
\caption{Comparison between experimental $\Z$ ($\bullet$,
Ref. \cite{BYS06}) and theoretical $\Z$ obtained as a sum of the direct
and resonant contributions, using the binding energy $|\eps _0|=0.3$~meV
(CH$_3$F), 25~meV (CH$_3$Cl), and 40~meV (CH$_3$Br). Vertical bars show
the energies of molecular fundamentals.}
\label{fig:exp}
\end{figure}

In Fig. \ref{fig:exp} we compare the beam-energy-averaged theoretical $\Z$
from Eq. (\ref{eq:Zeff_fin}) added to the direct contribution
$\Z ^{(\rm dir)}$, with measured $\Z$ for methyl halides \cite{BYS06,BGS03}.
Theoretical curves have been obtained using the binding energy of
$|\eps _0|=0.3$, 25, and 40 meV, for CH$_3$F, CH$_3$Cl and CH$_3$Br,
respectively.

Given the complexity of the problem and the fact that $\eps _0$
is the only free parameter in the calculation, the agreement between theory
and experiment in Fig. \ref{fig:exp} is remarkable.
In accord with Eq. (\ref{eq:Zeff_fin}), every vibrational mode
gives rise to a VFR, whose relative magnitude is determined by the
factor $g_\nu /k_\nu$ (since $\Gamma _\nu ^e/\Gamma _\nu \approx 1$).
On the positron longitudinal energy scale, the resonances are downshifted
from the mode energies by the positron binding energy and a further
12~meV due to the positron energy distribution.

Besides determining the resonance positions, the binding energy also affects
the overall magnitude of $\Zr$ via $\rho _{ep}\propto |\eps _0 |^{1/2}$
[Eq. (\ref{eq:rhoep})]. Hence, the smallness of $\Z$ in CH$_3$F in comparison
with those of CH$_3$Cl and CH$_3$Br is related to the weakness of its binding.
This is in turn related to the smaller dipole polarizability and higher
ionization potential of fluoromethane, which make it less attractive for
the positron.

Note that the infrared absorption strengths of the modes and the
corresponding elastic widths, $\Gamma _\nu ^e \sim \omega _\nu d_\nu ^2$, may
vary considerably from mode to mode. On the other hand, the contribution of
different modes to $\Z$ are similar, apart from energy shift and
$g_\nu /k_\nu$ factor. As a result, the energy dependence of $\Z$ has
little resemblance to the molecular infrared absorption spectra \cite{BYS06}.
The relation $\Gamma _\nu \approx \Gamma _\nu^e$ also means that the
contributions of the VFR are not sensitive to the exact values of the
elastic widths. Therefore, our use of the ``Born-dipole'' approximation in
the derivation of Eq.~(\ref{eq:Gam_eh}) is not expected to lead to sizeable
errors in $\Zr$.


In conclusion, we have presented a theory of positron annihilation
by capture into vibrational resonances of infrared-active modes. It agrees
well with measured $\Z$ for methyl halides and yields estimates of the
positron binding energies for these molecules.

This theory can also be used to investigate the contribution of
infrared-active-mode VFRs to $\Z$ in other small polyatomics that can
bind positrons. Such calculations will likely
underestimate the $\Z$, because the resonances associated with other
(non-dipole) modes may contribute just as much, as long as
their elastic widths are greater than the annihilation width.

In molecules where multiquantum vibrations are coupled by anharmonicity,
the number of VFRs populated by positron capture will be greatly increased,
leading to much higher $\Z$. However, the same coupling will also allow
the VFR to decay by positron emission to vibrationally excited states of
the molecule. This will increase the total resonance widths, thereby reducing
their individual contributions. Calculation of $\Z$ for molecules with IVR 
is the next big challenge for the theory.


The authors are grateful to C. M. Surko and J.~A.~Young for a most helpful
discussion of the positron energy distribution and valuable comments,
and to A.~V.~Korol for discussions.

\end{document}